\documentclass[aps,pre,twocolumn,superscriptaddress,showkeys,showpacs,amsmath,amssymb, longbibliography,nofootinbib ]{revtex4-1}

\usepackage{graphicx}
\usepackage{dcolumn}
\usepackage{bm}
\usepackage{hyperref}
\usepackage{xcolor}

\newcommand{\pin}[2]{#1_{\mathrm{#2}}}

\begin{document}
\title{Decoherence and information encoding in quantum reference frames}%
\author{Jan Tuziemski}%
\email{jan.tuziemski@fysik.su.se}
\altaffiliation[On leave from ]{Department of Applied Physics and Mathematics, Gdansk University of Technology}
\affiliation{Department of Physics, Stockholm University, AlbaNova University Center, Stockholm SE-106 91 Sweden}
\affiliation{Nordita, Royal Institute of Technology and Stockholm University,Roslagstullsbacken 23, SE-106 91 Stockholm, Sweden}

\date{\today}
\begin{abstract}
Reference frames are of special importance in physics. They are usually considered to be idealized entities. However, in most situations, e.g. in laboratories, physical processes are described within reference frames constituted by physical systems. As new technological developments make it possible to demonstrate quantum properties of complex objects an interesting conceptual problem arises: Could one use states of quantum systems to define reference frames? Recently such a framework has been introduced in [F. Giacomini, E. Castro-Ruiz, and Č. Brukner, Nat Commun 10, 494 (2019)]. One of its consequences is the fact that quantum correlations depend on a physical state of an observers reference frame. The aim of this work is to examine the dynamical aspect of this phenomena and show that the same is true for correlations established during an evolution of a composite systems. Therefore, decoherence process is also relative: For some observers the reduced evolution of subsystems is unitary, whereas for others it is not.  I also discuss implications of this results for modern developments of decoherence theory: Quantum Darwinism and spectrum broadcast structures.

\end{abstract}

\maketitle

\section{Introduction}
Description of physical phenomena is usually made within a reference frame. As a result, definition of certain physical quantities, such as states of physical systems, is relative and may change with a different choice of a reference frame. In most considerations reference frames are regarded as some idealized entities, whereas in practice, e.g. in laboratories, they are constituted by physical objects that are in states with well defined properties such as position or momentum. However, with the rapid development of the quantum technologies field new tests of the quantum superposition principle are made with ever heavier and more complex objects. Therefore, it is both important and interesting to consider extensions of reference frames to the quantum domain and a substantiation amount of work has been done in this direction, see e.g.  \cite{Aharonov1,Aharonov2,BartlettRMP,Palmer,Poulin2007,Rovelli1991quantum}. Here I follow a recent approach to that problem introduced in \cite{QRF} as it allows to associate reference frames with quantum states. As a consequence, a novel feature of this framework is that, in contrary to previous studies, it allows to consider reference frames with genuinely quantum features, e.g. being constituted by a state  in a superposition of distinct classical states. The main conclusion of \cite{QRF} is that correlations between subsystems of a composite quantum system are  relative to a quantum state of the observers reference frame. For instance, for one observer a system is in an entangled state with some other system, wheres for the other the state of the same system is product and hence uncorrelated.

Properties of quantum states of composite  systems are often used to explain emergence of classical behavior out of quantum laws. Let us consider a distinguished physical system prepared in a pure state that interacts with some degrees of freedom - the environment. Usually, due to the interaction, correlations are established  between the system and the environment. As a result, as the system evolves,  its state looses purity and some of its quantum features, such as coherences, may be suppressed.  This is the essence of decoherence theory: Initial quantum properties of the system become non-local features of the joint system-environment state. Moreover, recent developments of decoherence theory, i.e. quantum Darwinism \cite{ZurekNature} and spectrum broadcast structures \cite{myPRA,Korbicz2014}, put emphasis on the fact that during decoherence information about the system  is encoded in the environment.  Redundant information encoding allows to explain how information about decohered pointer states becomes locally available to many independent observers, what is an important feature of our everyday experience.

From the above discussion it is clear that dynamically established quantum correlations are essential to explain emergence of certain classical features out of quantum theory. On the other hand, if one adopts the framework of quantum reference frames \cite{QRF}, then quantum correlations are relative to a quantum state of a reference frame. Therefore one may ask whether decoherence and information encoding are also relative phenomena? 
Here I explore this question by investigating reduced dynamics of composite systems as seen from reference frames associated to different quantum states. An example of a transformation between two reference frames is shown, for which in one reference frame the reduced subsystem dynamics is unitary whereas in the second it is not. In one reference frame a subsystem undergoes decoherence and information about its pointer states in deposited in the other subsystem, whereas this not the case in the second reference frame. 

The paper is structured as follows: In Section \ref{sec:qrf} I briefly outline the quantum reference frame framework of \cite{QRF}. 
In  Sec. \ref{sec:model} the main results of the paper are presented: It is shown that a generalized Galilean transform between quantum reference frames leads to different description of reduced subsystems dynamics in the reference frames. Implication of this result for advanced forms of decoherence theory, quantum Darwinism and spectrum broadcast structures,  are discussed in Sec. \ref{sec:discuss}. Sec. \ref{sec:conclusions} concludes the paper, and an additional example is presented in the Appendix \ref{app:derG}.

\section{Quantum reference frames}
\label{sec:qrf}

\begin{figure}
	
	\includegraphics[scale=0.50]{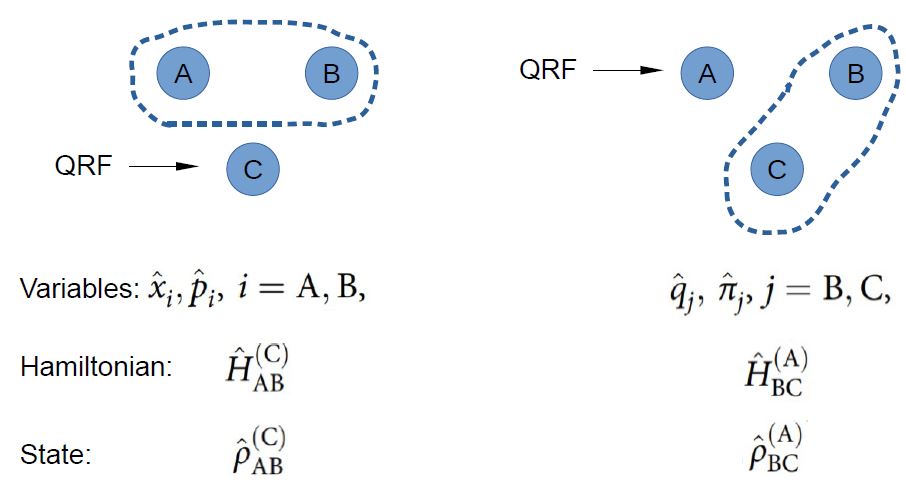}
	\caption{In the framework of \cite{QRF} a reference frame is associated to one part of the composite system: C (to the left) or A (to the right). In each reference frame there are canonical variables, state and Hamiltonian operator associated to the remaining subsystems. Quantities in quantum reference frames are related via a unitary operator $\hat{S}$, whose details depend on the chosen transformation.   \label{fig1:scheme}    }
	\centering
\end{figure}
Here we briefly review the main features of quantum reference frames formalism from Ref. \cite{QRF}. In the simplest, one-dimensional, scenario there are three quantum systems A, B, C, and one of them, say the system C, is chosen to be the quantum reference frame (see also Fig. \ref{fig1:scheme} ). Form the point of view of reference frame constituted by the state of C the quantum state of A and B is $\rho_{\mathrm{AB}}^{(\mathrm{C})} \in \mathcal{H}_{\mathrm{A}}^{(\mathrm{C})} \otimes \mathcal{H}_{\mathrm{B}}^{(\mathrm{C})}$. Evolution of this state is generated by a  Hamiltonian $\hat{H}_{\mathrm{AB}}^{(\mathrm{C})}$. We would like to see how this state and its dynamics is perceived from the perspective of subsystem A. In \cite{QRF} it was shown that a transformation between quantum reference frames of C and A is given by an appropriately chosen unitary transformation $\hat{S}: \mathcal{H}_{\mathrm{A}}^{(\mathrm{C})} \otimes \mathcal{H}_{\mathrm{B}}^{(\mathrm{C})} \rightarrow \mathcal{H}_{\mathrm{B}}^{(\mathrm{A})} \otimes \mathcal{H}_{\mathrm{C}}^{(\mathrm{A})}$.   The general structure of such a transformation is 
\begin{eqnarray}
\hat{S}=e^{-\frac{i}{\hbar} \hat{H}_{\mathrm{C}} t} \hat{\mathcal{P}}_{\mathrm{AC}}^{(i)} \Pi_{n} e^{\frac{i}{\hbar} \hat{f}_{\mathrm{A}}^{n}(t) \hat{O}_{\mathrm{B}}^{n}} e^{\frac{i}{\hbar} \hat{H}_{\mathrm{A}} t},
\end{eqnarray}
where $\hat{\mathcal{P}}_{\mathrm{AC}}: \mathcal{H}_{\mathrm{A}}^{(\mathrm{C})} \rightarrow \mathcal{H}_{\mathrm{C}}^{(\mathrm{A})}$ is a parity swap operator exchanging Hilbert spaces A with C, and  $\Pi_{n} e^{\frac{i}{\hbar} \hat{f}_{\mathrm{A}}^{n}(t) \hat{O}_{\mathrm{B}}^{n}}$ can be seen as
a quantum generalization of a transformation between classical reference frames, whose realization may in general require $n$ terms. For example a "coherent" translation   $\hat{U}_{X}=e^{\frac{1}{\hbar} \hat{x}_{\mathrm{A}} \hat{p}_{\mathrm{B}}}$ is a generalization of a quantum state transformation between two classical reference frames translated by a distance $X_0$: $\hat{U}_{X_0}=e^{\frac{1}{\hbar} X_0 \hat{p}_{\mathrm{B}}}$. Quantum state of subsystems AB transforms as 
\begin{eqnarray}
\label{eq:StateTrans}
\rho_{\mathrm{BC}}^{(\mathrm{A})}=\hat{S}\rho_{\mathrm{AB}}^{(\mathrm{C})} \hat{S}^{\dagger} \in \mathcal{H}_{\mathrm{B}}^{(\mathrm{A})} \otimes \mathcal{H}_{\mathrm{C}}^{(\mathrm{A})} ,
 \end{eqnarray}
 and its dynamics is governed by the standard von-Neumann equation
 \begin{eqnarray}
 \mathrm{i} \hbar \frac{d \hat{\rho}_{\mathrm{BC}}^{(\mathrm{A})}}{d t}=\left[\hat{H}_{\mathrm{BC}}^{(\mathrm{A})}, \hat{\rho}_{\mathrm{BC}}^{(\mathrm{A})}\right],
 \end{eqnarray}
with the transformed Hamiltonian 
\begin{eqnarray}
\label{eq:HamTrans}
\hat{H}_{\mathrm{BC}}^{(\mathrm{A})}=\hat{S} \hat{H}_{\mathrm{AB}}^{(\mathrm{C})} \hat{S}^{\dagger}+\mathrm{i} \hbar \frac{d \hat{S}}{d t} \hat{S}^{\dagger}.
\end{eqnarray}
As shown in Ref. \cite{QRF} in this framework correlations of quantum states become relative to the state of a reference frame. For example it may happen that, from the point of the reference frame associated to the state of subsystem C, subsystem B is in an entangled state with subsystem A, whereas in reference frame of A the state of subsystem B is pure and hence uncorrelated with subsystem C. Our aim here is to investigate how dynamics of subsystems is perceived in different reference frames. More precisely, we are interested in correlations that are established during subsystems evolution. Consider a situation, in which in A reference frame subsystem B becomes entangled with subsystem C, and this entanglement grows with time, whereas in  C reference frame subsystems A and B remain product during the evolution. Then in A reference frame the reduced evolution of subsystem B is non-unitary, B decoheres, and information about B could in principle be encoded in subsystem C. On the other hand in reference frame of C this is not the case. In the next Section it is shown that generalization of the Galilean transform provides such an example.     

\section{A case study: Generalized Galilean transform}
\label{sec:model}
The unitary operator $\hat{S}_\mathrm{b}$ that provides transformation between two reference frames related by the generalization of the Galilean transform is 
\begin{eqnarray}
\label{eq:Gtransf}
\hat{S}_{\mathrm{b}}=e^{-\frac{\mathrm{i}}{\hbar} \frac{\hat{\pi}_{\mathrm{C}}^{2}}{2 m_{\mathrm{C}}} t} \hat{\mathcal{P}}_{\mathrm{AC}}^{(\mathrm{v})} e^{\frac{\mathrm{i}}{\hbar} \frac{\hat{p}_{\mathrm{A}}}{m_{\mathrm{A}}} \hat{G}_{\mathrm{B}}}e^{\frac{\mathrm{i}}{\hbar} \frac{\hat{p}_{\mathrm{A}}^{2}}{2 m_{\mathrm{A}}} t},
\end{eqnarray} 
where
$
\hat{G}_{\mathrm{B}}=\hat{p}_{\mathrm{B}} t-m_{\mathrm{B}} \hat{x}_{\mathrm{B}},
$
and and $\hat{\mathcal{P}}_{\mathrm{AC}}^{(\mathrm{v})}$ maps velocity of A to the opposite velocity of C:
$
 \hat{\mathcal{P}}_{\mathrm{AC}}^{(\mathrm{v})}=\hat{\mathcal{P}}_{\mathrm{AC}} \exp \left(\frac{\mathrm{i}}{\hbar} \log \sqrt{\frac{m_{\mathrm{C}}}{m_{\mathrm{A}}}}\left(\hat{x}_{\mathrm{A}} \hat{p}_{\mathrm{A}}+\hat{p}_{\mathrm{A}} \hat{x}_{\mathrm{A}}\right)\right).
$
Let us assume that in the reference frame of observer C systems A and B are free particles. Their Hamiltonian is thus 
\begin{eqnarray}
\label{eq:ABHam}
\hat{H}_{\mathrm{AB}}^{(\mathrm{C})}=\frac{\hat{p}_{\mathrm{A}}^{2}}{2 m_{\mathrm{A}}}+\frac{\hat{p}_{\mathrm{B}}^{2}}{2 m_{\mathrm{B}}},
\end{eqnarray}
and we assume that the particles are initially uncorrelated 
$
\left|\Psi^{(\mathrm{C})}_0\right\rangle_{\mathrm{AB}} = \left|\varphi_0\right\rangle_{\mathrm{A}} \otimes \left|\psi_0\right\rangle_{\mathrm{B}}
\ 
$
According to the transformation law Eq. (\ref{eq:StateTrans}), the evolution of the initial state in reference frame of A is   
$
\left|\Psi^{(\mathrm{A})}(t)\right\rangle_{\mathrm{CB}} = 
\hat{S}_{\mathrm{b}} e^{-\frac{i}{\hbar} \hat{H}_{\mathrm{AB}}^{\mathrm{(C)}} t }\left|\Psi^{(\mathrm{C})}_0 \right\rangle_{\mathrm{AB}}.
$
 The state after this transformation is
\begin{eqnarray}
\label{eq:AGalistateAppm}
&&\left|\psi^{(\mathrm{A})}(t)\right\rangle_{\mathrm{BC}}= 
 e^{-\frac{i}{\hbar} \hat{H}_{\mathrm{BC}}^{\mathrm{(A)}} t }  e^{-\frac{\mathrm{i}}{\hbar} \frac{\hat{\pi}_{\mathrm{C}}}{m_{\mathrm{C}}} \hat{G}_{\mathrm{B}}}  \left|\Psi^{(\mathrm{A})}_0 \right\rangle_{\mathrm{BC}}. \nonumber \\
\end{eqnarray} 
For simplicity we assume that $\pin{m}{A}=\pin{m}{C}$.  
The initial state in the transformed frame is still a product one
\begin{eqnarray}
&&\left|\Psi^{(\mathrm{A})}_0\right\rangle_{\mathrm{BC}} =
\int \int  d \pi_{\mathrm{B}} d \pi_{\mathrm{C}}   \psi_{0}\left(\pi_{\mathrm{B}}\right) \left|\pin{\pi}{B}\right\rangle_{\mathrm{B}} \otimes \phi_0(-\pin{\pi}{C})\left|\pin{\pi}{C}\right\rangle_{\mathrm{C}}, \nonumber \\
\end{eqnarray}
where the minus in $\phi_0(-\pin{\pi}{C})$ indicates that subsystem C  evolves with opposite velocity to that of A.
To investigate dynamics of $\left|\Psi^{(\mathrm{A})}(t)\right\rangle_{\mathrm{BC}}$ it is convenient to rewrite the Galilean operator using the Baker - Campbell - Hausdorff formula 
\begin{eqnarray}
\label{eq:GalBCHm}
&&e^{-\frac{i}{h} \frac{\hat{\pi}_{\mathrm{C}}}{m_{\mathrm{C}}} \hat{G}_{\mathrm{B}}} = \int d \pin{\pi}{C}  e^{-\frac{i}{h} \frac{\pi_{\mathrm{C}}}{m_{\mathrm{C}}} \hat{G}_{\mathrm{B}}} \otimes \left| \pin{\pi}{C}  \right \rangle \left \langle \pin{\pi}{C} \right|_{\mathrm{C}}  =  \\ \nonumber &&\int d \pin{\pi}{C}  e^{-\frac{i}{h} \frac{\pin{m}{B}}{m_{\mathrm{C}}} \frac{\pi^2_{\mathrm{C}}}{2 \pin{m}{C}}  t }  e^{-\frac{i}{h} \frac{\pi_{\mathrm{C}}}{m_{\mathrm{C}}} \pin{\hat{\pi}}{B} t } e^{\frac{i}{h} \frac{ \pin{m}{B}}{m_{\mathrm{C}}} \pi_{\mathrm{C}} \pin{\hat{q}}{B} } \otimes \left| \pin{\pi}{C}  \right \rangle \left \langle \pin{\pi}{C} \right|_{\mathrm{C}} .
\end{eqnarray}
 The form of Eq. (\ref{eq:GalBCHm}) suggests that the evolution of subsystem B is controlled by the momentum of subsystem C. One can also show that evolution of subsystem C is controlled by subsystem B. For the sake of argument we choose the initial B state to be approximately a "sharp" position state such that initially  $\pin{\hat{q}}{B} \left| \psi_0 \right \rangle_{\mathrm{B}} \approx 0$, and therefore having large coherences in momentum basis (a more general case of a cat state of subsystem B is discussed in Appendix \ref{app:derG}). Therefore we can write
\begin{eqnarray}
\label{eq:AGalistate}
&&\left|\Psi^{(\mathrm{A})}(t)\right\rangle_{\mathrm{BC}} = 
 e^{-\frac{i}{\hbar} \hat{H}_{\mathrm{BC}}^{\mathrm{(A)}} t } \int  d \pi_{\mathrm{B}}  \psi_{0}\left(\pi_{\mathrm{B}}\right) \left|\pin{\pi}{B}\right\rangle_{\mathrm{B}} \otimes  \hat{U}_{\pin{\pi}{B}}(t) \left| \phi_0 \right \rangle_{\mathrm{C}}, \nonumber \\
\end{eqnarray}
where
\begin{eqnarray}
 &&\hat{U}_{\pin{\pi}{B}}(t) \left| \phi_0 \right \rangle_{\mathrm{C}} \equiv \left| \phi_{\pin{\pi}{B}}(t) \right \rangle_C  \nonumber = \\&& \int d \pi_{\mathrm{b}}  e^{-\frac{i}{h} \frac{\pi^2_{\mathrm{C}}}{2 m}  t }  e^{-\frac{i}{h} \frac{\pi_{\mathrm{B}}}{\pin{m}{C}} \pin{\pi}{C} t } \phi_0(-\pin{\pi}{C})\left|\pin{\pi}{C}\right\rangle_{\mathrm{C}}, \nonumber \\&&  
\end{eqnarray}
 what demonstrates that momentum of subsystem B controls evolution of subsystem C and $\hat{H}_{\mathrm{BC}}^{(\mathrm{A})}$  is given by Eq. (\ref{eq:HamTrans}), where the transformation $\hat{S}$ is Eq. (\ref{eq:Gtransf}). In the considered case the functional form of Hamiltonian is preserved, i.e.  $\hat{H}_{\mathrm{BC}}^{(\mathrm{A})}=\frac{\hat{\pi}_{\mathrm{B}}^{2}}{2 m_{\mathrm{B}}}+\frac{\hat{\pi}_{\mathrm{C}}^{2}}{2 m_{\mathrm{C}}}$, so in A reference Hamiltonians of subsystems C and B also  are also those of free particles. The state (\ref{eq:AGalistate}) is, in general, an entangled state of subsystems C and B. The source of this entanglement is the generalized Galilean term Eq. (\ref{eq:Gtransf}), in which momentum of subsystem A is promoted to an operator. As a result, the situation here is somehow similar to that considered in the case of a novel mechanism of gravitational decoherence \cite{PikovskiM,Pikovski2,myGrawitacja}, in which frequency of internal degrees of freedom effectively  becomes an operator. Note also that in the present case division into the system and the environment is not as clear as in most models of decoherence and one could in principle treat the subsystem B as the environment. We choose to study decoherence of subsystem B as it is perceived both in reference frame of A and C. 
 
\begin{figure}
	\includegraphics[scale=0.40]{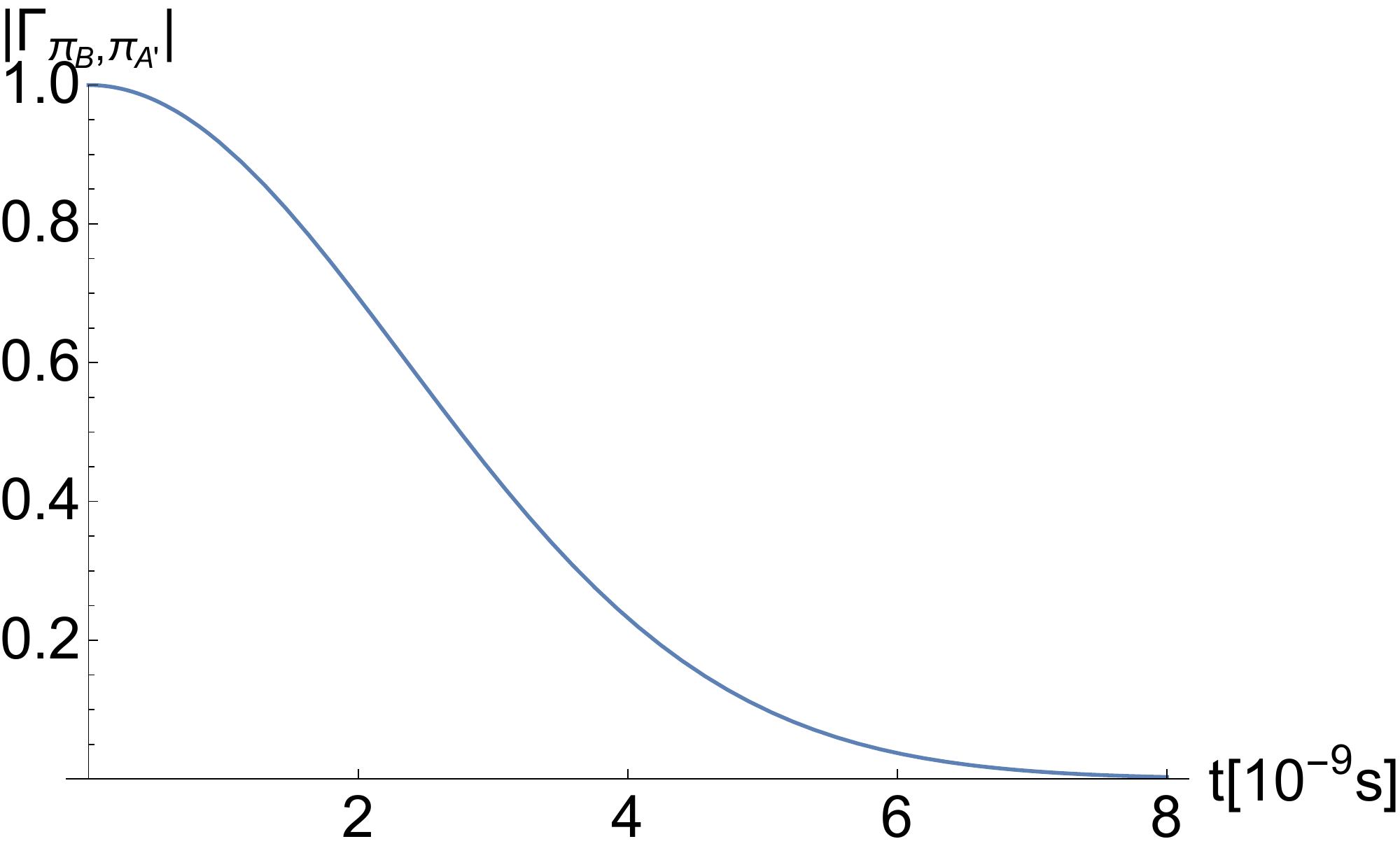}
	\caption{Modulus of decoherence factor Eq. (\ref{eq:decf}) as a function time. The parameters are: $\frac{\pin{\pi}{B}-\pin{\pi'}{B}}{\Delta_{\gamma_0}} = 2 \times 10^{-6}  $, $\pin{m}{C}=10^{-17} kg$. As explained in the main text, for pure initial states modulus of decoherence factor is equal to generalized overlap, thus in the considered case this plot also shows time dependence of generalized overlap Eq. (\ref{eq:genover}) for the same choice of parameters. \label{fig2:decf}    }
	\centering
\end{figure}

 Let us now turn attention to the reduced state of subsystem B. In C reference frame the state of subsystem B is initially pure and remains pure during the evolution.  On the other hand, Eq. (\ref{eq:AGalistate}) indicates that this in not the case in A reference frame as one finds  
 \begin{eqnarray}
 \label{eq:AredB}
 && \pin{\hat{\rho}}{B}(t) = Tr_{\mathrm{C}} \left[\left|\psi^{(\mathrm{A})} (t)\right\rangle \left \langle\psi^{(\mathrm{A})}(t)\right|_\mathrm{BC}\right] = \\ && \nonumber  \int \int d \pin{\pi}{B} d \pin{\pi'}{B}  \Gamma_{ \pin{\pi}{B},\pin{\pi'}{B}}(t) \psi_0(\pin{\pi}{B})\psi_0^*(\pin{\pi'}{B}) \left| \pin{\pi}{B} \right \rangle \left \langle \pin{\pi'}{B} \right|_{\mathrm{B}},
 \end{eqnarray}
 where (assuming that $m_{\mathrm{A}}=m_{\mathrm{C}})$
 \begin{eqnarray}
 \label{eq:AdecF}
\Gamma_{ \pin{\pi'}{B},\pin{\pi}{B}}(t) \equiv  &&\left \langle \phi_{\pin{\pi}{B}}(t) \right| \left. \phi_{\pin{\pi'}{B}}(t) \right \rangle  = \\ &&\int d \pin{\pi}{C} e^{-\frac{i}{h} \frac{\pi_{\mathrm{c}}}{m_{\mathrm{C}}} \left(\pin{\pi}{B} - \pin{\pi'}{B} \right)  t } \left| \phi_0(-\pin{\pi}{C}) \right|^2. \nonumber
\end{eqnarray}
For simplicity we choose $\phi_0(\pin{\pi}{C})$ to be a Gaussian $\phi_0(\pin{\pi}{C})=\left(\Delta_{\gamma_0}^2 \pi \right)^{-1/4} e^{- \left(\pin{p}{C} -\gamma_0 \right)^2/(2\Delta_{\gamma_0}^2)} $, then decay of coherences is a Gaussian one
\begin{eqnarray}
\label{eq:decf}
\Gamma_{ \pin{\pi}{B},\pin{\pi'}{B}}(t) = e^{-\left(\frac{t}{\tau_{\Delta \pin{\pi}{B}}}\right)^2}
e^{-\frac{i \gamma_0 }{\hbar \pin{m}{C} }  \left(\pin{\pi}{B} -\pin{\pi'}{B} \right)  t },
\end{eqnarray}
where
\begin{eqnarray}
\label{eq:dectime}
{\tau_{\Delta \pin{\pi}{B}}} \equiv  \frac{2 \hbar \pin{m}{C}}{ \left(\pin{\pi}{B} -\pin{\pi'}{B}\right) \Delta_{\gamma_0}}.
\end{eqnarray}  
The above result implies that in the A reference frame the state of B subsystem will decohere in momentum.  Timescale of this process is determined by the ratio $\pin{m}{C}/\Delta_{\gamma_0}$ so for well-localized states of massive systems there is no decoherence since $\tau_{\Delta \pin{\pi}{B}} \rightarrow \infty$, and one recovers the standard Galilean transform. Here we are interested in an opposite regime. A sample plot of modulus of decoherence factor is shown in Fig. \ref{fig2:decf}.

The above analysis shows that, as time passes, in the A reference frame subsystems B and C become entangled, and this in turn leads to decoherence of the state of subsystem B  (for pure states the only source of decoherence is entanglement). One finds that, in momentum basis, off-diagonal elements of subsystem B density matrix are suppressed. As have been shown in recent developments of decoherence theory, usually the subsystem causing decoherence (in our case the subsystem C) carries some information about decohering system. In order to check whether this is the case we investigate properties of the reduced state of subsystem C, which is 
\begin{eqnarray}
\pin{\hat{\rho}}{C}(t) = \int \pin{\pi}{B} \left| \psi_{0} \left(\pi_{\mathrm{B}}\right) \right|^2 \left| \phi_{\pin{\pi}{B}}(t) \right \rangle  \left \langle \phi_{\pin{\pi}{B}}(t) \right|_\mathrm{C}.  
\end{eqnarray}
The structure of this state suggests that information about subsystem B momentum is encoded in subsystem C. Therefore, one needs to check how distinguishable are states of C corresponding to two different momenta $\pin{\pi}{B},\pin{\pi'}{B}$ . A useful measure of distinguishability of quantum states is generalized overlap, which for two density matrices $\hat{\rho}_{\mathrm{\pi_B}}, \hat{\rho}_{\mathrm{\pi'_B}}$ reads \cite{FG1999}
\begin{eqnarray}
B_{\mathrm{\pi_B},\pi_{\mathrm{B}}'} \equiv Tr \sqrt{\sqrt{\hat{\rho}_{\pi_{\mathrm{B}}'} } \hat{\rho}_{\pi_{\mathrm{B}}} \sqrt{\hat{\rho}_{\pi_{\mathrm{B}}' }} }.
\end{eqnarray}
Generalized overlap allows to quantify how distiguishable the two states are:  $B_{\mathrm{\pi_B},\pi_{\mathrm{B}}'}=0$ indicates that they are perfectly distiguishable, whereas for $B_{\mathrm{\pi_B},\pi_{\mathrm{B}}'}=1$ they cannot be distinguished. Moreover, it allows to construct measurements optimally discriminating these states. Generalized overlap is used e.g. as a tool to check whether a multi-partite state retains a specific structure known as  spectrum broadcast structure  \cite{myPRA,MironowiczPRL}. 
 For pure states generalized overlap reduces to the standard state overlap. In the case discussed here $\hat{\rho}_{\mathrm{\pi_B}}(t)\equiv \left| \phi_{\pin{\pi}{B}}(t) \right \rangle  \left \langle \phi_{\pin{\pi}{B}}(t) \right|_\mathrm{C}$ and $\hat{\rho}_{\mathrm{\pi'_B}}(t)\equiv \left| \phi_{\pin{\pi'}{B}}(t) \right \rangle  \left \langle \phi_{\pin{\pi'}{B}}(t) \right|_\mathrm{C}$ so $B_{\mathrm{\pi_B},\pi_{\mathrm{B}}'}  = \left| \left \langle \phi_{\pin{\pi'}{B}}(t) \right. \left| \phi_{\pin{\pi}{B}}(t) \right \rangle  \right|$. Therefore, in the considered case decoherence of subsystem B automatically implies information encoding in subsystem C and the time dependence of this process is 
 \begin{eqnarray}
 \label{eq:genover}
 B_{\pin{\pi}{B}, \pin{\pi'}{B}}= \left|\Gamma_{\pin{\pi}{B}, \pin{\pi'}{B}}\right| =  e^{-\left(\pin{\pi}{B} -\pin{\pi'}{B} \right)^2 \left(\frac{ t}{\tau_{dec} } \right)^2 }. 
 \end{eqnarray}       

From Eqs. (\ref{eq:decf}) and (\ref{eq:genover}) one can conclude that in A reference frame subsystem B decoheres and information about momentum is encoded in subsystem C. From the point of subsystem A subsystem B neither decoheres nor is information about its momentum encoded anywhere (state of B remains pure during the evolution). This the main result of the present work: Decoherence and information encoding is relative to the quantum state of the observers reference frame.

\section{Advanced forms of decoherence in quantum reference frames}   
 \label{sec:discuss}
In the previous section it was demonstrated that if the physical states of reference frames are incorporated into description of the considered physical problem as in Ref. \cite{QRF}, then decoherence and encoding of information into the system causing decoherence are relative phenomena that depend on the state of the observers reference frame.

Let us now discuss implications of the above findings for recently studied advanced forms of decoherence: Quantum Darwinism and spectrum broadcast structures (SBS) \cite{ZurekNature,myPRA}, which recently have become a vivid theoretical (see e.g. \cite{DarTheor1,DarTheor2,DarTheor3,DarTheor4,Roszak1,Roszak2,PostquantumSBS}) and experimental (\cite{DarExp1, DarExp2, DarExp3}) area of research. Those frameworks aim to explain some aspects of quantum-to-classical transition problem. Decoherence accounts for the fact that states of complex (macroscopic) objects are  rarely found in some state being a superposition of distinct classical states. In addition to that quantum Darwinism and SBS explain how it is possible for multiple observers to independently retrieve the same information about the state of decohering system, and do not disturb it nor results of measurements of other observers. These are properties of classical states. A multipartite quantum state describing the system and some available to observers fragment of the environment can exhibit these properties since, usually, the  environment causing decoherence consists of numerous degrees of freedom, and each of them carries some information about decohereing systems hence this information is redundantly encoded in the environment (it may also be the case that a single degree of freedom is not enough to retrieve information about the system but a group of them is, such groups are refereed to as macrofractions \cite{myPRA}).
Quantum Darwinism as well as SBS provide tools in order to check whether in a considered situation quantum state will exhibit such classical features. In general both approaches are not equivalent as conditions of SBS are stricter than those of quantum Darwinism. A multipartite state that admits SBS form fulfills also conditions of quantum Darwinism, whereas the converse statement is not true \cite{Lee2019}. Therefore here we focus on SBS and the conclusions reached here hold also for quantum Darwinism. The state admits spectrum broadcast structure form if it can be written as    

\begin{eqnarray}
&&\hat{\sigma}_{S B S}=\sum_{i} p_{i}|i\rangle\langle i| \otimes \hat{\rho}_{i}^{1} \ldots \otimes \hat{\rho}_{i}^{N} \\&& \nonumber
\hat{\rho}_{i}^{k} \perp \hat{\rho}_{i^{\prime}}^{k} \;  \textrm{for every} \;
 i^{\prime} \neq i \; \textrm{and} \; k=1, \ldots, N.
\end{eqnarray}
The basis $\left\{ | i \rangle\right\}$ constitutes the so-called pointer states of the central system to which it decoheres, $p_i$ are initial pointer probabilities, $k$ enumerates the environments, and $\hat{\rho}_{i}^{k}$ are some
states of the observed parts of the environment which have mutually orthogonal supports for different pointer index $i$ so that they are
perfectly distinguishable with the help of a single measurement. 

Let us consider a trivial extension of the model considered in the previous Section to the case, in which subsystems A and C consist of two degrees of freedom and the transformation (\ref{eq:Gtransf}) concerns the joint momentum $\pin{\hat{p}}{A_1A_2} = \pin{\hat{p}}{A_1} \otimes \hat{I} + \hat{I} \otimes  \pin{\hat{p}}{A_2}$, and similarly for subsystem C. Then if in the A reference frame one part of subsystem C is not observed the joint state will be
 \begin{eqnarray}
\label{eq:SBSrel}
&& \pin{\hat{\rho}}{\mathrm{BC_1}}(t) = Tr_{\mathrm{C_2}} \left[\left|\psi^{(\mathrm{A})} (t)\right\rangle \left \langle\psi^{(\mathrm{A})}(t)\right|_\mathrm{BC_1C_2}\right] = \\ &&\nonumber  \int \int d \pin{\pi}{B} d \pin{\pi'}{B}  \Gamma_{ \pin{\pi}{B},\pin{\pi'}{B}}(t) \psi_0(\pin{\pi}{B})\psi_0^*(\pin{\pi'}{B})  \left| \pin{\pi}{B} \right \rangle \left \langle \pin{\pi'}{B} \right|_{\mathrm{B}} \otimes  \nonumber \\&&\left| \phi_{\pin{\pi}{B}}(t) \right \rangle \left \langle  \phi_{\pin{\pi}{B}}(t) \right|_{\mathrm{C_1}},
\end{eqnarray}
where $\Gamma_{ \pin{\pi}{B},\pin{\pi'}{B}}(t)$ is given by Eq. \ref{eq:AdecF}. For the same choice of the initial state for subsystems $\mathrm{C_1C_2}$ as in the previous Section one can see that the structure of the above state, after decoherence time, will be in a good approximation a SBS one (for discussion on SBS in continuous variables setting see e.g. \cite{Tuziemski2015b,myGrawitacja,myQED}). On the other hand, in C reference frame there will be no decoherence or SBS formation even if one subsystem of A is traced out. Therefore, formation of SBS is relative to the choice of reference frame and SBS and quantum Darwinism frameworks, in order to correctly capture the mentioned aspects of quantum-to-classical transition, need to assume existence of classical reference frames between different observers. As has been discussed in the previous Section,  in the limit of large masses and small coherences of the quantum states that define quantum reference frames one recovers classical reference frames. Therefore, for observers using complex objects to constitute their reference frames this requirement is usually satisfied.
\section{Conclusions}
\label{sec:conclusions}
In this paper I investigated reduced dynamics of subsystems as seen form different quantum reference frames, using the framework introduced in \cite{QRF}. Using the generalized Galiliean transform I demonstrated that both decoherence and information encoding about decohering system in the environment are relative to a quantum state constituting the reference frame. I discussed implications of this results for quantum Darwinism and SBS, arguing that, to correctly capture some features of quantum-to-classical transition, those frameworks need to assume existence of classical reference frames.

\textit{Note added-} While this work was being completed a preprint appeared \cite{LeMH}, in which a similar conclusion is reached, namely that SBS are relative to quantum reference frames. However, there the static case is studied, i.e. the transformation between reference frames does not involve momentum. Therefore, the results reported here are complementary to those of Ref. \cite{LeMH}.   
\section*{Acknowledgment}
\label{ack}
Helpful remarks from J. K. Korbicz are acknowledged. This work was supported  by  the European Research Council under grant 742104.

\bibliographystyle{unsrt}
\bibliography{polaron}

\begin{thebibliography}{10}

\bibitem{Aharonov1}
Yakir Aharonov and Leonard Susskind.
\newblock Charge superselection rule.
\newblock {\em Physical Review}, 155(5):1428, 1967.

\bibitem{Aharonov2}
Yakir Aharonov and Leonard Susskind.
\newblock Observability of the sign change of spinors under 2 $\pi$ rotations.
\newblock {\em Physical Review}, 158(5):1237, 1967.

\bibitem{BartlettRMP}
Stephen~D Bartlett, Terry Rudolph, and Robert~W Spekkens.
\newblock Reference frames, superselection rules, and quantum information.
\newblock {\em Reviews of Modern Physics}, 79(2):555, 2007.

\bibitem{Palmer}
Matthew~C Palmer, Florian Girelli, and Stephen~D Bartlett.
\newblock Changing quantum reference frames.
\newblock {\em Physical Review A}, 89(5):052121, 2014.

\bibitem{Poulin2007}
David Poulin and Jon Yard.
\newblock Dynamics of a quantum reference frame.
\newblock {\em New Journal of Physics}, 9(5):156, 2007.

\bibitem{Rovelli1991quantum}
Carlo Rovelli.
\newblock Quantum reference systems.
\newblock {\em Classical and Quantum Gravity}, 8(2):317, 1991.

\bibitem{QRF}
Flaminia {Giacomini}, Esteban {Castro-Ruiz}, and {\v{C}}aslav {Brukner}.
\newblock {Quantum mechanics and the covariance of physical laws in quantum
  reference frames}.
\newblock {\em Nature Communications}, 10:494, January 2019.

\bibitem{ZurekNature}
Wojciech~Hubert Zurek.
\newblock Quantum darwinism.
\newblock {\em Nat Phys}, 5:181--188, March 2009.

\bibitem{myPRA}
R.~Horodecki, J.~K. Korbicz, and P.~Horodecki.
\newblock Quantum origins of objectivity.
\newblock {\em Phys. Rev. A}, 91:032122, Mar 2015.

\bibitem{Korbicz2014}
J.~K. Korbicz, P.~Horodecki, and R.~Horodecki.
\newblock Objectivity in a noisy photonic environment through quantum state
  information broadcasting.
\newblock {\em Phys. Rev. Lett.}, 112:120402, Mar 2014.

\bibitem{PikovskiM}
Igor {Pikovski}, Magdalena {Zych}, Fabio {Costa}, and {\v{C}}aslav {Brukner}.
\newblock {Universal decoherence due to gravitational time dilation}.
\newblock {\em Nature Physics}, 11(8):668--672, August 2015.

\bibitem{Pikovski2}
Igor Pikovski, Magdalena Zych, Fabio Costa, and Časlav Brukner.
\newblock Time dilation in quantum systems and decoherence.
\newblock {\em New Journal of Physics}, 19(2):025011, 2017.

\bibitem{myGrawitacja}
J.~K. Korbicz and J.~Tuziemski.
\newblock Information transfer during the universal gravitational decoherence.
\newblock {\em General Relativity and Gravitation}, 49(12):152, Nov 2017.

\bibitem{FG1999}
C.~A. Fuchs and J.~van~de Graaf.
\newblock Cryptographic distinguishability measures for quantum-mechanical
  states.
\newblock {\em IEEE Transactions on Information Theory}, 45(4):1216--1227, May
  1999.

\bibitem{MironowiczPRL}
P.~Mironowicz, J.~K. Korbicz, and P.~Horodecki.
\newblock Monitoring of the process of system information broadcasting in time.
\newblock {\em Phys. Rev. Lett.}, 118:150501, Apr 2017.

\bibitem{DarTheor1}
Guillermo {Garc{\'\i}a-P{\'e}rez}, Dario~A. {Chisholm}, Matteo A.~C. {Rossi},
  G.~Massimo {Palma}, and Sabrina {Maniscalco}.
\newblock {Decoherence without entanglement and quantum Darwinism}.
\newblock {\em Physical Review Research}, 2(1):012061, March 2020.

\bibitem{DarTheor2}
Steve {Campbell}, Bar{\i}{\textcommabelow s} {{\c{c}}akmak}, {\"O}zg{\"u}r~E.
  {M{\"u}stecapl{\i}o{\v{g}}lu}, Mauro {Paternostro}, and Bassano {Vacchini}.
\newblock {Collisional unfolding of quantum Darwinism}.
\newblock {\em \pra}, 99(4):042103, April 2019.

\bibitem{DarTheor3}
Salvatore {Lorenzo}, Mauro {Paternostro}, and G.~Massimo {Palma}.
\newblock {Anti-Zeno-based dynamical control of the unfolding of quantum
  Darwinism}.
\newblock {\em Physical Review Research}, 2(1):013164, February 2020.

\bibitem{DarTheor4}
Salvatore {Lorenzo}, Mauro {Paternostro}, and G.~{Massimo Palma}.
\newblock {Reading a qubit quantum state with a quantum meter: time unfolding
  of quantum Darwinism and quantum information flux}.
\newblock {\em arXiv e-prints}, page arXiv:2001.11558, January 2020.

\bibitem{Roszak1}
Katarzyna {Roszak} and Jaros{\l}aw~K. {Korbicz}.
\newblock {Entanglement and objectivity in pure dephasing models}.
\newblock {\em \pra}, 100(6):062127, December 2019.

\bibitem{Roszak2}
Katarzyna {Roszak} and Jaros{\l}aw~K. {Korbicz}.
\newblock {A glimpse of objectivity in bipartite systems for non-entangling
  pure dephasing evolutions}.
\newblock {\em arXiv e-prints}, page arXiv:1911.05558, November 2019.

\bibitem{PostquantumSBS}
Carlo~Maria {Scandolo}, Roberto {Salazar}, Jaros{\l}aw~K. {Korbicz}, and
  Pawe{\l} {Horodecki}.
\newblock {Is it possible to be objective in every physical theory?}
\newblock {\em arXiv e-prints}, page arXiv:1805.12126, May 2018.

\bibitem{DarExp1}
Mario~A. Ciampini, Giorgia Pinna, Paolo Mataloni, and Mauro Paternostro.
\newblock Experimental signature of quantum darwinism in photonic cluster
  states.
\newblock {\em Phys. Rev. A}, 98:020101, Aug 2018.

\bibitem{DarExp2}
T.~K. Unden, D.~Louzon, M.~Zwolak, W.~H. Zurek, and F.~Jelezko.
\newblock Revealing the emergence of classicality using nitrogen-vacancy
  centers.
\newblock {\em Phys. Rev. Lett.}, 123:140402, Oct 2019.

\bibitem{DarExp3}
Ming-Cheng Chen, Han-Sen Zhong, Yuan Li, Dian Wu, Xi-Lin Wang, Li~Li, Nai-Le
  Liu, Chao-Yang Lu, and Jian-Wei Pan.
\newblock Emergence of classical objectivity of quantum darwinism in a photonic
  quantum simulator.
\newblock {\em Science Bulletin}, 64(9):580 -- 585, 2019.

\bibitem{Lee2019}
Thao~P. Le and Alexandra Olaya-Castro.
\newblock Strong quantum darwinism and strong independence are equivalent to
  spectrum broadcast structure.
\newblock {\em Phys. Rev. Lett.}, 122:010403, Jan 2019.

\bibitem{Tuziemski2015b}
J.~Tuziemski and J.~K. Korbicz.
\newblock Dynamical objectivity in quantum brownian motion.
\newblock {\em EPL}, 112(4):40008, 2015.

\bibitem{myQED}
J.~Tuziemski, P.~Witas, and J.~K. Korbicz.
\newblock Redundant information encoding in qed during decoherence.
\newblock {\em Phys. Rev. A}, 97:012110, Jan 2018.

\bibitem{LeMH}
Thao~P. {Le}, Piotr {Mironowicz}, and Pawe{\l} {Horodecki}.
\newblock {Blurred quantum Darwinism across quantum reference frames}.
\newblock {\em arXiv e-prints}, page arXiv:2006.06364, June 2020.

\end{thebibliography}
\begin{widetext}
\appendix
\section{Derivation of decoherence factor for a cat state of subsystem B}
\label{app:derG}
Here we present a detailed derivation of the decoherence factor in the case of the generalized Galilean transform between reference frames, for a more general choice of subsystems B state. 
\\
We choose the initial state to be a product one
\begin{eqnarray}
\left|\Psi^{(\mathrm{A})}_0\right\rangle_{\mathrm{BC}} = \hat{\mathcal{P}}_{\mathrm{AC}}^{(\mathrm{v})} \left|\Psi^{(\mathrm{C})}_0\right\rangle_{\mathrm{AB}} 
   = \int \int  d \pi_{\mathrm{B}} d \pi_{\mathrm{C}}   \psi_{0}\left(\pi_{\mathrm{B}}\right) \left|\pin{\pi}{B}\right\rangle_{\mathrm{B}} \otimes \phi_0(-\pin{\pi}{C})\left|\pin{\pi}{C}\right\rangle_{\mathrm{C}},
\end{eqnarray}
where $\phi_0\left(\pin{p}{C} \right)=  \left(\Delta_{\gamma_0}\pi \right)^{-1/4} e^{- \left(\pin{p}{C} -\gamma_0 \right)^2/(2\Delta_{\gamma_0}^2)}$ is a Gaussian and  and $ \psi_0\left( \pin{p}{B} \right) = \sqrt{N}^{-1} \left(\Delta_{\gamma_0}\pi \right)^{-1/4} \left( e^{- \left(\pin{p}{B} -\beta \right)^2/(2\Delta_{\beta_0}^2)}  + e^{- \left(\pin{p}{B} -\beta' \right)^2/(2\Delta_{\beta_0}^2)}  \right),
$ is a cat state in momentum space. 
Using Eq. (\ref{eq:GalBCHm}) one finds
\begin{eqnarray}
\left|\Psi^{(\mathrm{A})}(t)\right\rangle_{\mathrm{BC}} =&& e^{-\frac{i}{\hbar} \hat{H}_{\mathrm{BC}}^{\mathrm{(A)}} t } \int \int  d \pi_{\mathrm{B}} d \pi_{\mathrm{C}}   \psi_{0}\left(\pin{\pi}{B}- \frac{\pin{m}{B}}{m_{\mathrm{C}}} \pi_{\mathrm{C}} \right) \left|\pin{\pi}{B}\right\rangle_{\mathrm{B}} \otimes  e^{-\frac{i}{h} \frac{\pi^2_{\mathrm{C}}}{2 m_{\mathrm{C}}}  t }  e^{-\frac{i}{h} \frac{\pi_{\mathrm{B}}}{m_{\mathrm{C}}} \pin{\pi}{C} t } \phi_0(-\pin{\pi}{C})\left|\pin{\pi}{C}\right\rangle_{\mathrm{C}}
\end{eqnarray}
The initial wavefunction of B is superposition of Gaussian states, therefore one can write
\begin{eqnarray}
\psi \left(\pin{\pi}{B}- \frac{\pin{m}{B}}{m_{\mathrm{C}}} \pi_{\mathrm{C}}  \right) = \sum_{\tilde{\beta} \in \beta,\beta'} e^{- \left(\pin{\pi}{B} -\tilde{\beta} \right)^2/(2\Delta_{\beta_0}^2)} e^{\left(\pin{\pi}{B} - \tilde{\beta}\right) \frac{\pin{m}{B} }{m_{\mathrm{C}}\Delta_{\beta_0}^2} \pi_{\mathrm{C}}  }e^{- \left( \frac{m_{\mathrm{B}}} {m_{\mathrm{C}} \Delta_{\beta_0}} \right)^2 \frac{\pi_{\mathrm{C}}^2}{2} }. 
\end{eqnarray}
In this way one the state can be rewritten is a form similar to that of Eq. (\ref{eq:AGalistate}) in the main text 
\begin{eqnarray}
\left|\Psi^{(\mathrm{A})}(t)\right\rangle_{\mathrm{BC}} =&& e^{-\frac{i}{\hbar} \hat{H}_{\mathrm{BC}}^{\mathrm{(A)}} t } \sum_{\tilde{\beta} \in \beta,\beta'} \int  d \psi_{\tilde{\beta}}\left(\pin{\pi}{B}\right) \left|\pin{\pi}{B}\right\rangle_{\mathrm{B}} \otimes \left| \tilde{\phi}_{\pin{\pi}{B},\tilde{\beta}} \right \rangle_{\mathrm{C}}
\end{eqnarray}
where
\begin{eqnarray}
&&\psi_{\tilde{\beta}}\left(\pin{\pi}{B}\right) \equiv e^{- \left(\pin{\pi}{B} -\tilde{\beta} \right)^2/(2\Delta_{\beta_0}^2)} \\&& 
\left| \tilde{\phi}_{\pin{\pi}{B},\tilde{\beta}} \right \rangle \equiv \int d \pi_{\mathrm{C}} e^{\left[-\frac{i}{h} \frac{\pi_{\mathrm{B}}}{m_{\mathrm{C}}}  t+\left(\pin{\pi}{B} - \tilde{\beta}\right) \frac{\pin{m}{B} }{m_{\mathrm{C}}\Delta_{\beta_0}^2}\right] \pi_{\mathrm{C}}  }e^{ -\left[ \frac{i}{h m_{\mathrm{C}} }   t+\left(\frac{m_{\mathrm{B}}} {m_{\mathrm{C}} \Delta_{\beta_0}} \right)^2  \right]\frac{\pi_{\mathrm{C}}^2}{2}   } \phi_0(-\pin{\pi}{C})\left|\pin{\pi}{C}\right\rangle_{\mathrm{C}}
\end{eqnarray}

The reduced state of the B subsystem is thus
\begin{eqnarray}
Tr_\mathrm{C} \left[\left|\psi^{(\mathrm{A})} (t)\right\rangle \left \langle\psi^{(\mathrm{A})}(t)\right|_{\mathrm{CB}}\right] = N^{-1} \int \int d \pin{\pi}{B} d \pin{\pi'}{B}  \sum_{\tilde{\beta} \in \beta,\beta'} \sum_{\tilde{\beta}' \in \beta,\beta' } \tilde{\Gamma}_{ \pin{\pi}{B},\pin{\pi'}{B},\tilde{\beta},\tilde{\beta}'}(t)  \psi_{\tilde{\beta}} \left(\pin{\pi}{B} \right) \psi_{\tilde{\beta}'} \left(\pin{\pi'}{B} \right),
\end{eqnarray}
where
\begin{eqnarray}
\tilde{\Gamma}_{ \pin{\pi}{B},\pin{\pi'}{B},\beta,\beta' } \equiv \left \langle  \tilde{\phi}_{\pin{\pi'}{B},\beta'} \right| \left.  \tilde{\phi}_{\pin{\pi}{B},\beta} \right \rangle = \int d \pi_{\mathrm{C}} e^{-\frac{i}{h} \left( \pin{\pi}{B} - \pin{\pi'}{B}\right) \frac{\pi_{\mathrm{C}}}{m_{\mathrm{C}}} t } e^{\left(\pin{\pi}{B} + \pin{\pi'}{B} -\beta-\beta'\right) \frac{\pin{m}{B} }{m_{\mathrm{C}}\Delta_{\beta_0}^2} \pi_{\mathrm{C}}  }e^{- \left( \frac{m_{\mathrm{B}}} {m_{\mathrm{C}} \Delta_{\beta_0}} \right)^2 \pi_{\mathrm{C}}^2 } \left|\varphi_0 \left(-\pin{\pi}{C}\right)\right|^2.
\end{eqnarray}
To simplify expression we will choose masses to be equal $m_{\mathrm{A}}=m_{\mathrm{B}}=m_{\mathrm{C}}=m$.

It is convenient to split the result of the above integration into two parts, in which one depends on $\beta, \beta'$ and the other doesn't. Explicitly one has:
\begin{eqnarray}
&&\tilde{\Gamma}_{ \pin{\pi}{B},\pin{\pi'}{B},\beta,\beta' }(t)= \Gamma_{ \pin{\pi}{B},\pin{\pi'}{B},}(t) \Gamma_{ \pin{\pi}{B},\pin{\pi'}{B},\beta,\beta'}(t),
\end{eqnarray}
where
\begin{eqnarray}
\label{eq:decfactg}
&& \Gamma_{ \pin{\pi}{B},\pin{\pi'}{B},}(t)= \Gamma_{ \pin{\pi}{B},\pin{\pi'}{B},}(0) e^{\frac{\Delta_{\gamma_0}^2\Delta_{\beta_0}^2}{4\left(\Delta_{\gamma_0}^2+\Delta_{\beta_0}^2\right)} \left\{ \frac{-\left(\pin{\pi}{B}-\pin{\pi'}{B}\right)^2t^2}{\hbar^2 m^2} - 2 \frac{i}{\hbar} \frac{\left(\pin{\pi}{B}-\pin{\pi'}{B}\right)t}{m}\left[ 2\frac{\gamma_0}{\Delta_{\gamma_0}^2}   +  \frac{\left( \pin{\pi}{B} + \pin{\pi'}{B}\right)}{\Delta_{\beta_0}^2}\right] \right\} } \\
&&
\Gamma_{ \pin{\pi}{B},\pin{\pi'}{B} }(0) = \sqrt{\frac{\Delta_{\gamma_0}^2}{\Delta_{\beta_0}^2+\Delta_{\gamma_0}^2}} e^{\frac{\Delta_{\gamma_0}^2\Delta_{\beta_0}^2}{4\left(\Delta_{\gamma_0}^2+\Delta_{\beta_0}^2\right)} \left\{ 4\frac{\gamma_0\left( \pin{\pi}{B} + \pin{\pi'}{B}\right) }{\Delta_{\gamma_0}^2  \Delta_{\beta_0}^2} +  \frac{ \left( \pin{\pi}{B} + \pin{\pi'}{B}\right)^2}{\Delta^2_{\beta_0} }-4\frac{ \gamma_0^2}{\Delta_{\gamma_0}^2\Delta_{\beta_0}^2}  \right\} } \\
&& \Gamma_{ \pin{\pi}{B},\pin{\pi'}{B},\beta_0,\beta_0'}(t)  = \Gamma_{ \pin{\pi}{B},\pin{\pi'}{B},\beta_0,\beta_0'}(0) e^{\frac{i\Delta_{\gamma_0}^2\left(\pin{\pi}{B}-\pin{\pi'}{B}\right) \left(\beta+\beta'\right)t}{2 \hbar m \left(\Delta_{\gamma_0}^2+\Delta_{\beta_0}^2\right)} }\\
&& 
\Gamma_{ \pin{\pi}{B},\pin{\pi'}{B},\beta,\beta' }(0) =  e^{\frac{\Delta_{\gamma_0}^2\Delta_{\beta_0}^2}{4\left(\Delta_{\gamma_0}^2+\Delta_{\beta_0}^2\right)} \left\{ -4\frac{\gamma_0\left( \beta+\beta'\right)}{\Delta_{\gamma_0}^2\Delta_{\beta_0} ^2}   + \left(\frac{ \beta+\beta'}{\Delta_{\beta_0}^2}  \right)^2 -2 \frac{ \beta+\beta'}{\Delta_{\beta_0}^2}\left( \pin{\pi}{B}+\pin{\pi'}{B}\right)   \right\} }. \\
\end{eqnarray} 
In this case there two factors contributing to the loss of coherence. The first is that the initial state is no longer the product one. Therefore, even for $t=0$, one has  $\left| \tilde{\Gamma}_{ \pin{\pi}{B},\pin{\pi'}{B},\beta,\beta' }(0)\right| < 1 $ since the initial state is an entangled one. In addition to that entanglement between B and C subsystems grows with time as for the simpler example considered in the main text. In fact, the leading term casing dynamical suppresion of B coherences is only a slight modification of that presented in the main text as
\begin{eqnarray}
\left| \tilde{\Gamma}_{ \pin{\pi}{B},\pin{\pi'}{B},\beta,\beta' }(0)\right| \sim e^{-\left(\frac{t}{\tilde{\tau}_{\Delta \pin{\pi}{B}}}\right)^2},
\end{eqnarray}
where (cf. Eq. \ref{eq:dectime} of the main text)
\begin{eqnarray}
{\tilde{\tau}_{\Delta \pin{\pi}{B}}} \equiv  \frac{2 \hbar m \sqrt{\Delta_{\gamma_0}^2+\Delta_{\beta_0}^2}}{ \left(\pin{\pi}{B} -\pin{\pi'}{B}\right) \Delta_{\gamma_0}\Delta_{\beta_0}}. 
\end{eqnarray}  
Regarding the information content of the subsystem B, as we again deal with pure initial states, decreasing decoherence factor implies that distinguishability of states of subsystem C corresponding to different momenta $\pin{\pi}{B},\pin{\pi'}{B}$ increases with time.

\end{widetext}

\end{document}